\documentclass[twocolumn, trackchanges]{aastex62}

\newcommand{\tess}{{\it TESS}}
\newcommand{\kep}{{\it Kepler}}

\newcommand{\plato}{{\it PLATO}}

\newcommand{\mstar}{$M_{\star}$}

\newcommand{\rsun}{{R$_\odot$}}
\newcommand{\msun}{{M$_\odot$}}
\newcommand{\mearth}{$M_{\oplus}$}
\newcommand{\rearth}{$R_{\oplus}$}
\newcommand{\chisquare}{$\chi^2$}
\newcommand{\q}{$Q_{\ast}$}
\newcommand{\qcrit}{$Q_{\ast, c}$}

\usepackage{hyperref}

\usepackage{ tipa }
\usepackage{ amssymb }
\usepackage{ amsmath }

\newcommand{\chicago}{Department of Astronomy and Astrophysics, University of
Chicago, 5640 S. Ellis Ave, Chicago, IL 60637, USA}
\newcommand{\sagan}{Sagan Fellow}

\shorttitle{\textit{TESS} Observations of the \textit{Kepler} Field}
\shortauthors{Christ et al.}

\begin{document}

\title{Observations of the \textit{Kepler} Field with \textit{TESS}: Predictions for Planet Yield and Observable Features}

%% Note that the corresponding author command and emails has to come
%% before everything else. Also place all the emails in the \email
%% command instead of using multiple \email calls.
\correspondingauthor{Callista~N.~Christ}
\email{callistac@uchicago.edu}

\author[0000-0002-9346-4183]{Callista~N.~Christ}
\affiliation{\chicago}

\author[0000-0001-7516-8308]{Benjamin~T.~Montet}
\altaffiliation{\sagan}
\affiliation{\chicago}

\author[0000-0003-3750-0183]{Daniel~C.~Fabrycky}
\affiliation{\chicago}

%% The \author command can take an optional ORCID.

%% Note that RNAAS manuscripts DO NOT have abstracts.

\begin{abstract}
     We examine the ability of the Transiting Exoplanet Survey Satellite (\tess) to detect and improve our understanding of planetary systems in the \kep\ field. By modeling the expected transits of all confirmed and candidate planets detected by \kep\ as expected to be observed by \tess, we provide a probabilistic forecast of the detection of each \kep\ planet in \tess\ data. We find that 
      \tess\ has a greater than 50\% chance of detecting 260 of these planets at the $3\sigma$ level in one sector of observations and an additional 120 planets in two sectors. Most of these are large planets in short orbits around their host stars, although a small number of rocky planets are 
      expected to be recovered. Most of these systems have only one known transiting planet; in only $\sim 5$ percent of known multiply-transiting systems do we anticipate more than one planet to be recovered.
      When these planets are recovered, we expect \tess\ to be a powerful tool to characterizing transit timing variations. Using Kepler-88 (KOI-142) as an example, we show that \tess\ will improve measurements of planet-star mass ratios and orbital parameters, and significantly reduce the transit timing uncertainty in future years. Since \tess\ will be most sensitive to hot Jupiters, we research whether \tess\ will be able to detect tidal orbital decay in these systems. We find two confirmed planetary systems (Kepler-2\,b and Kepler-13\,b) and five candidate systems that will be good candidates to detect tidal decay. 
\end{abstract}


\section{Introduction}
The \kep\ spacecraft \citep{Borucki10} is a groundbreaking instrument that has detected thousands of exoplanets, including several that are Earth-sized and lie in the habitable zone of their host stars \citep{Rowe15, Mullally15, Coughlin16, Thompson18}.
It has also altered the way that we think about the formation and structure of planetary systems \citep{Lissauer11, Fang12}.
Although \kep\ has been a valuable tool thus far in exoplanetary studies, most \kep\ stars are too faint for detailed follow-up, such as obtaining precise RV measurements to determine the planets' masses \citep{Ricker16}.

Nevertheless, if a system has more than one planet, we can utilize the system's transit timing variations \citep[TTVs,][]{Agol05, Holman05} to teach us more about the system. 
TTVs occur in multi-planet systems due to gravitational interactions between planets and can be visible in transit timing data since they force planets off of a strictly Keplerian orbit. 
TTVs can be used to not only confirm planetary systems but also to measure system mass ratios and orbital parameters \citep{Fabrycky13, Huber13, Nespral16}. 
Furthermore, TTVs allow us to significantly improve our characterization of planetary systems by deriving the physical parameters of the systems, such as the star/planets' absolute mass, eccentricity, and inclinations, especially when combined with other data \citep{Agol05, Montet13, Almenara18}. 

Since the mass measurements from TTV signals strongly depend on orbital parameters such as period ratio and eccentricity and since for many systems we do not have enough data from \kep\ to measure these precisely \citep{Fabrycky13}, we have either large uncertainties or only upper limits in mass measurements \citep{Nesvorny12}.
Additionally, many TTVs are still degenerate after four years since the period of many TTV systems are comparable in length or longer than the \kep\ observing baseline \citep{Holczer16, Mazeh13}. 
This means that in order to obtain more precise measurements of TTVs, orbital parameters, and mass measurements, we need to increase the \kep\ four year baseline by continuing to observe these systems.

This is where \tess, the Transiting Exoplanet Survey Satellite \citep{Ricker16}, comes into play. 
\tess\ was launched in April 2018, and will survey 80\% of the sky to catalog the nearest and brightest stars in our local neighborhood \citep{Ricker16}.
This will make \tess\ planets some of the best characterized planets that exist since consistent follow-up observations will be easily performed in the future.
Since \tess\ will be performing a nearly all-sky survey, it will re-observe the \kep\ field. 
Thus, \tess\ will extend the amount of time we have observed these systems from four to ten years which will allow us to examine the long-term dynamical effects that exist in planetary systems.

This analysis will be critical in multiple ways. First, we may discover additional planets which did not transit in the \kep\ era but will transit when \tess\ observes the system (e.g. dynamical perturbations may allow previously non-transiting planets to transit, and/or the planets' periods were longer than the  \kep\ four-year baseline). Additionally, more data on these systems will allow us to better constrain the systems' planetary parameters.  

In this paper we discuss the procedure that utilizes previously obtained \kep\ data with soon to be obtained \tess\ data to improve planetary parameters' measurements. 
We demonstrate how we can use \kep\ data in conjunction with \tess's predicted transit times to improve our measurements of various systems and to allow us to explore other effects such as tidal decay of hot Jupiters. We show that this will be achievable in a single sector of \tess\ observations, but also that with additional data, either during the primary or an extended mission, \tess's ability to characterize the small planets originally detected by \kep\ increases by a factor of several.

The rest of this paper is organized as follows.
In Section 2, we discuss our method for discovering what types of planets \tess\ is sensitive to. 
In Section 3, we determine how well we can improve our measurements of masses and eccentricities with \tess\ data by analyzing a best-case scenario system, KOI-142.
In Section 4, we examine the detectable planets from Section 2 and determine how many of these systems we have a strong chance of observing tidal orbital decay with \tess.
In Section 5, we offer conclusions and look to the future.

\section{\tess\ Sensitivity to Kepler's Transiting Planets}
\subsection{Finding Probabilities of Detection}
\label{sec:methods}
Our goal in this section is to determine how many and what types of planets \tess\ will detect in the \kep\ field, thus increasing the observation baseline for these systems from four to ten years. 
This increased baseline with \tess\ will improve our measurements of systems and will allow us to better and more accurately test theories on planetary formation and migration.

We use the stellar and planetary properties from  \cite{Mathur17} for systems in the NASA Exoplanet Archive to determine the types of planets \tess\ will be sensitive to. 
We consider planets labeled as ``confirmed'' or ``candidate'' and neglect any systems identified as false positives.
We contaminate the transit depths from \cite{Thompson18} with \tess's  contamination ratios \citep{Stassun17} to find the transit depths that \tess\ is expected to observe for each system.

To find the total uncertainty expected in \tess\ photometry for each system, we apply the projected noise estimate given in Fig. 14 of \cite{Sullivan15} and retrieve the total noise for each observation, given a \tess\ apparent magnitude \citep{Stassun17}.

Given the total noise, the calculated \tess\ transit depth, the \kep\ transit duration, orbital period, and an exposure time of 30 minutes, we compute the expected signal-to-noise ratios (SNR) for each planet.
In the above calculations, we compute the SNR for both the case that \tess\ will observe the \kep\ field for one or two sectors: the length of observation for any given star will depend on the exact pointing of the telescope in 2019, but much of the field may be observed for two sectors.
The exact pointing of the telescope in the Northern hemisphere will not be determined until 2019, due to uncertainties in the future orbit and the details of Earth-moon crossings across the detector light path, which may necessitate changes to the exact sector positioning.

To convert these SNRs to probabilities of detection we follow \cite{Christiansen15}, assuming the probability of detection is a function of the observed SNR, following a logistic function centered at a $7.1\sigma$ detection threshold\footnote{See for example the \tess\ data release notes which describe the \tess\ transit search pipeline.}.
This idealized scenario was not achieved for \kep, but may be for \tess, depending on the noise characteristics in real data from that instrument.
We also consider detection at the $3\sigma$ threshold, 
as a lower significance might be acceptable for the characterization of known planets, rather than the discovery of new planets.
We retrieve probabilities of detection for each planet, weighted by the relative likelihoods of observing $N$ transits given the planet's orbital period and the \tess\ observing baseline. We repeat this analysis under the scenario that each planet is observed for either one or two sectors.
Thus, we found two probabilities of detection per planet: one probability given that the star will be observed for one sector, and another that the star will be observed for two sectors. The results are shown in Table \ref{tab:tess_sensitivity}.

\subsection{Results}
\label{sec:sensitivity}

\begin{figure*}[!tbh]
  \centering
  \begin{minipage}[b]{0.48\textwidth}
    \includegraphics[width=\textwidth]{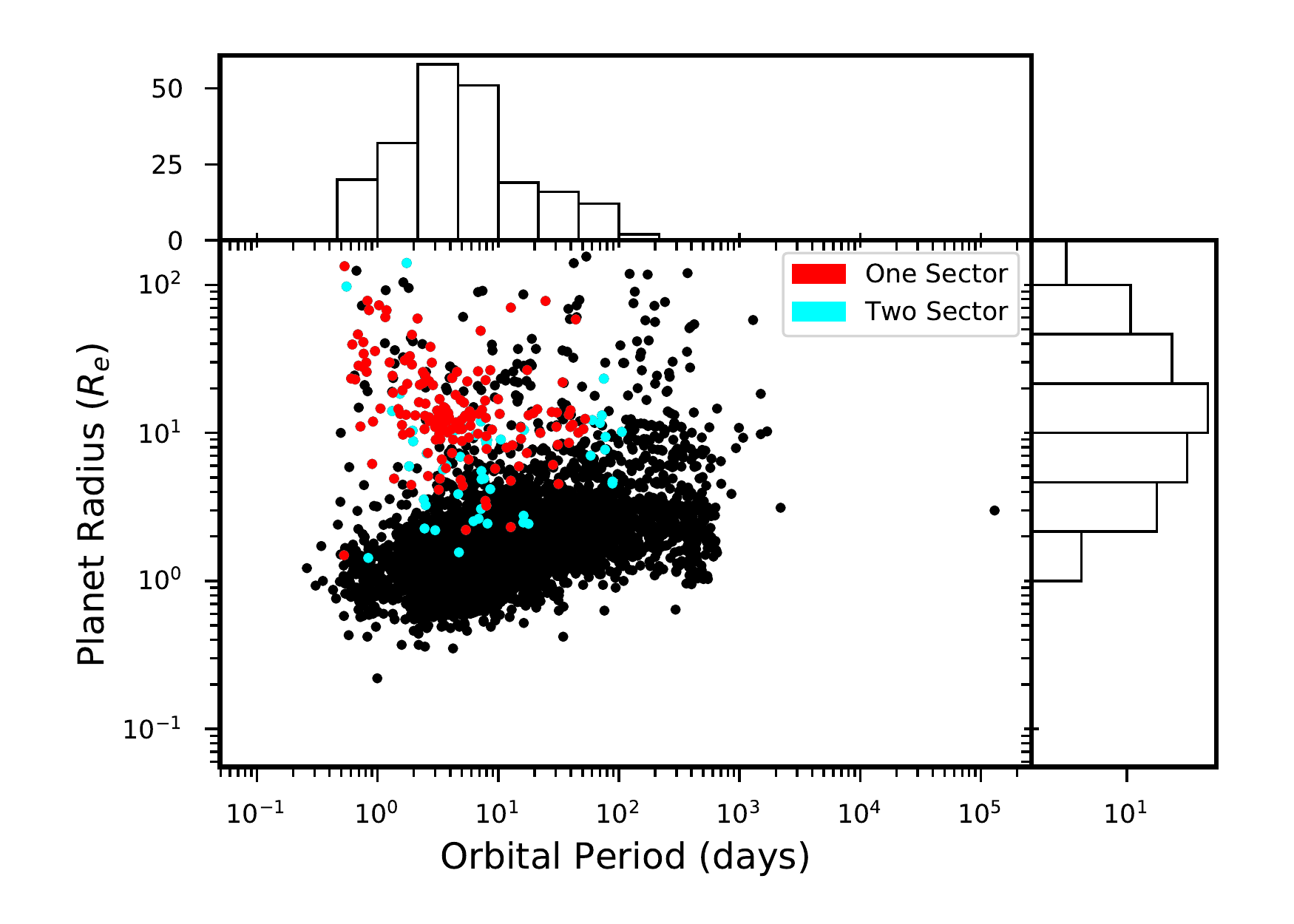}
  \end{minipage}
  \hfill
  \begin{minipage}[b]{0.48\textwidth}
    \includegraphics[width=\textwidth]{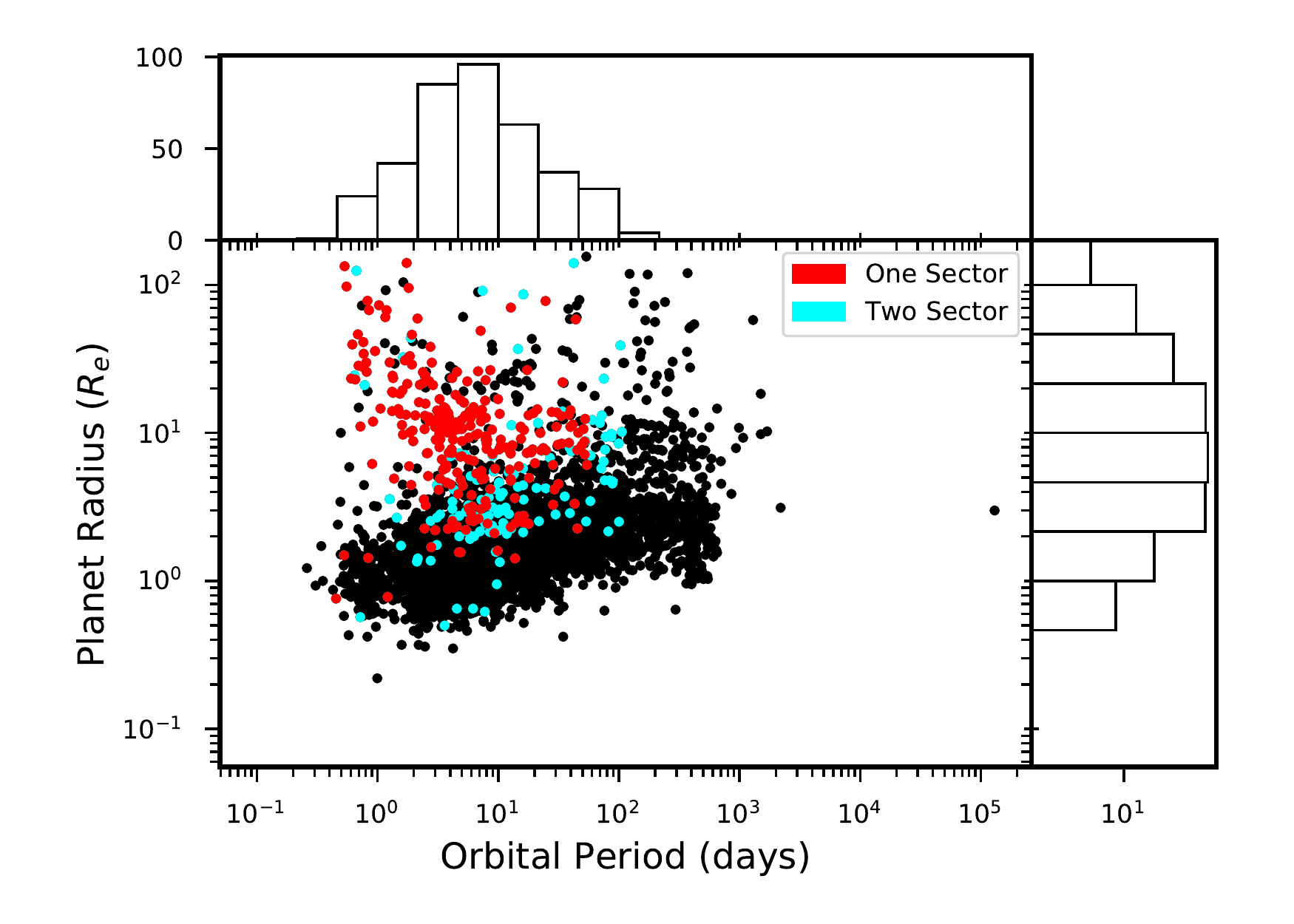}
  \end{minipage}
  \caption{(Left) A radius versus period plot of all confirmed and candidate planets detectable by \kep. Given a detection threshold of $7.1\sigma$, the red data points are planets that have a probability greater than 50\% of being detected by \tess\ in one sector (27.4 days). The blue data points are the additional planets that have a probability greater than 50\% of being detected by \tess\ if it observes the \kep\ field in two sectors (54.8 days). The histograms on the x and y-axis represent the spread in orbital period and planet radius, respectively. (Right) The same plot as the left except with a $3\sigma$ detection threshold. Out of the 4,456 \kep\ planets in total, \tess\ will only be able to recover a small portion of these signals; nevertheless, \tess\ is expected to detect a large number of hot Jupiter planets and will thus be able to improve our measurements of these systems.}
  \label{fig:RvsP}
\end{figure*}

Figure \ref{fig:RvsP} shows the radius and period of all confirmed and candidate planets discovered by \kep, highlighting the systems that we predict have a greater than 50\% chance of being detected by \tess.
At the $7.1\sigma$ level, we find 81 (114) confirmed planets and 80 (96) candidate planets that will be detected by \tess\ in one (two) sector(s) of observations.
At the $3\sigma$ level, there are 154 (232) confirmed and 106 (148) candidate planets that are detectable in one (two) sector(s). 
In total, we expect \tess\ is likely to recover 260 (380) of these signals originally detected by \kep\ in one (two) sectors. 
A \tess\ mission that is positioned to spend two sectors of observing time covering the \kep\ field rather than one will likely detect more than 120 additional planet signals from the \kep\ field alone. Most of the additional detections will be planets smaller than Neptune. 

Additional observations of the \kep\ field during an extended mission will continue to contribute to the characterization of these small planets (Table 1).
However, the timing of these additional observations does not significantly affect the future yield of these planets, as the most significant present limitation is the signal to noise ratio of the individual planets in \tess\ data. To maximize the detections and ability to characterize known, transiting planets, the number of which that can be detected are as shown in Table 1, the detailed scheduling of these observations is unimportant relative to the number of sectors spent observing the \kep\ field, which should be maximized. Here, the largest gain would be for the detection and possible characterization of Sub-Neptune planets, which in most cases can not be detected in 1-2 sectors of observations around \kep\ targets. These observations will be important however, to detect the possible long-term precession of these systems by distant, non-transiting perturbers similar to those observed by \textit{K2} (Hamann et al. in prep).

\begin{deluxetable*}{ccccc}
\tablecaption{Cumulative number of \kep\ planets detectable in \tess\ data at the $3\sigma$ level after a given number of sectors of observations. A longer extended mission will lead to a considerable growth in the number of small planets recovered.}
\tablewidth{0pt}
\tablehead{
  \colhead{Number of sectors} &
  \colhead{Candidate Planets} & 
  \colhead{Confirmed Planets} &
  \colhead{Candidate Sub-Neptunes} &
  \colhead{Confirmed Sub-Neptunes}
}
\startdata
1 & 106 & 154 & 7 & 36 \\
2 & 148 & 232 & 18 & 89 \\
3 & 181 & 302 & 29 & 131 \\
4 & 206 & 351 & 37 & 169 \\
5 & 229 & 396 & 50 & 206 \\
6 & 251 & 454 & 60 & 252 
\enddata
\label{tab:newplanets}
\end{deluxetable*}

The full list of \kep\ objects of interest and their detection probabilities with \tess\ is given in Table 2.
A majority of the planets that are detectable at either the $3\sigma$ or $7.1\sigma$ level are large planets in short orbits around their host star: most of the \kep\ planets that \tess\ will detect are hot Jupiters. 
Nevertheless, it is still important to note that smaller planets can be detected as long as their host star is bright and small in size. 
For example, the planet candidate KOI-06635.01 orbits a 14.353 magnitude star with a radius of 0.41 \rsun\ with an orbital period of 0.5274 days. Despite being $1.5$ \rearth\ in size, we project this planet will be detectable.

\begin{deluxetable*}{lcccccccc}
%\tabletypesize{\scriptsize}
\tablecaption{A list of expected signals and computed probabilities of detection at the $3\sigma$ level for both one and two sectors of observation. The systems are ordered by KOI name.}
\tablewidth{0pt}
\tablehead{
  \colhead{KOI} &
  \colhead{\kep\ Name} & 
  \colhead{\tess\ Mag.\tablenotemark{1}} &
  \colhead{\tess\ \tablenotemark{2}} &
  \colhead{Total Noise \tablenotemark{3}} &
  \colhead{SNR} &
  \colhead{SNR} &
  \colhead{Probability} &
  \colhead{Probability} \\
  \colhead{} &
  \colhead{} & 
  \colhead{} &
  \colhead{Transit Depth} &
  \colhead{} &
  \colhead{1-Sector} &
  \colhead{2-Sector} &
  \colhead{1-Sector} &
  \colhead{2-Sector} \\
  \colhead{} & 
  \colhead{} &
  \colhead{} &
  \colhead{(ppm)} &
  \colhead{(ppm)} &
  \colhead{} &
  \colhead{} &
  \colhead{} &
  \colhead{} 
}
\startdata
K00001.01 & Kepler-1\,b & 10.79 & 13810 & 451 & 190.30 & 269.14 & 1.000 & 1.000  \\
K00002.01 & Kepler-2\,b & 10.00 & 6596 & 301 & 215.49 & 304.80 & 1.000 & 1.000 \\
K00003.01 & Kepler-3\,b & 8.44 & 4239 & 152 & 142.99 & 202.38 & 1.000 & 1.000  \\
K00004.01 & & 10.98 & 1169 & 500 & 14.32 & 20.25 & 1.000 & 1.000 \\
K00007.01 & Kepler-4\,b & 11.67 & 663 & 727 & 7.51 & 10.63 & 1.000 & 1.000 \\
K00010.01 & Kepler-8\,b & 13.08 & 9088 & 1636 & 39.16 & 55.40 & 1.000 & 1.000 \\
K00012.01 & Kepler-448\,b  & 10.89 & 6824 & 475 & 67.54 & 96.79 & 1.000 & 1.000 \\
K00013.01 & Kepler-13\,b & 9.57 & 4548 & 244 & 184.68 & 261.21 & 1.000 & 1.000 \\
K00017.01 & Kepler-6\,b  & 12.62 & 8430 & 1248 & 52.68 & 74.53 & 1.000 & 1.000 \\
K00018.01 & Kepler-5\,b & 12.70 & 6757 & 1309 & 43.33 & 61.29 & 1.000 & 1.000 \\
\enddata
\tablenotetext{1}{Taken from \cite{Stassun17}.}
\tablenotetext{2}{Calculated using Kepler and TESS contamination ratios from \cite{Brown11} and \cite{Stassun17}, respectively.}
\tablenotetext{3}{Obtained by inputting \tess\ apparent magnitudes into the function from \cite{Sullivan15}'s Fig. 14.}
\tablecomments{This table will be published in its entirety in machine-readable format on the journal website. 
A portion is reproduced here as a guide for formatting. A version is also available in the source materials
for this manuscript on the arXiv.}
\label{tab:tess_sensitivity}
\end{deluxetable*}

\begin{figure*}[!tbh]
  \centering
  \begin{minipage}[b]{0.48\textwidth}
    \includegraphics[width=\textwidth]{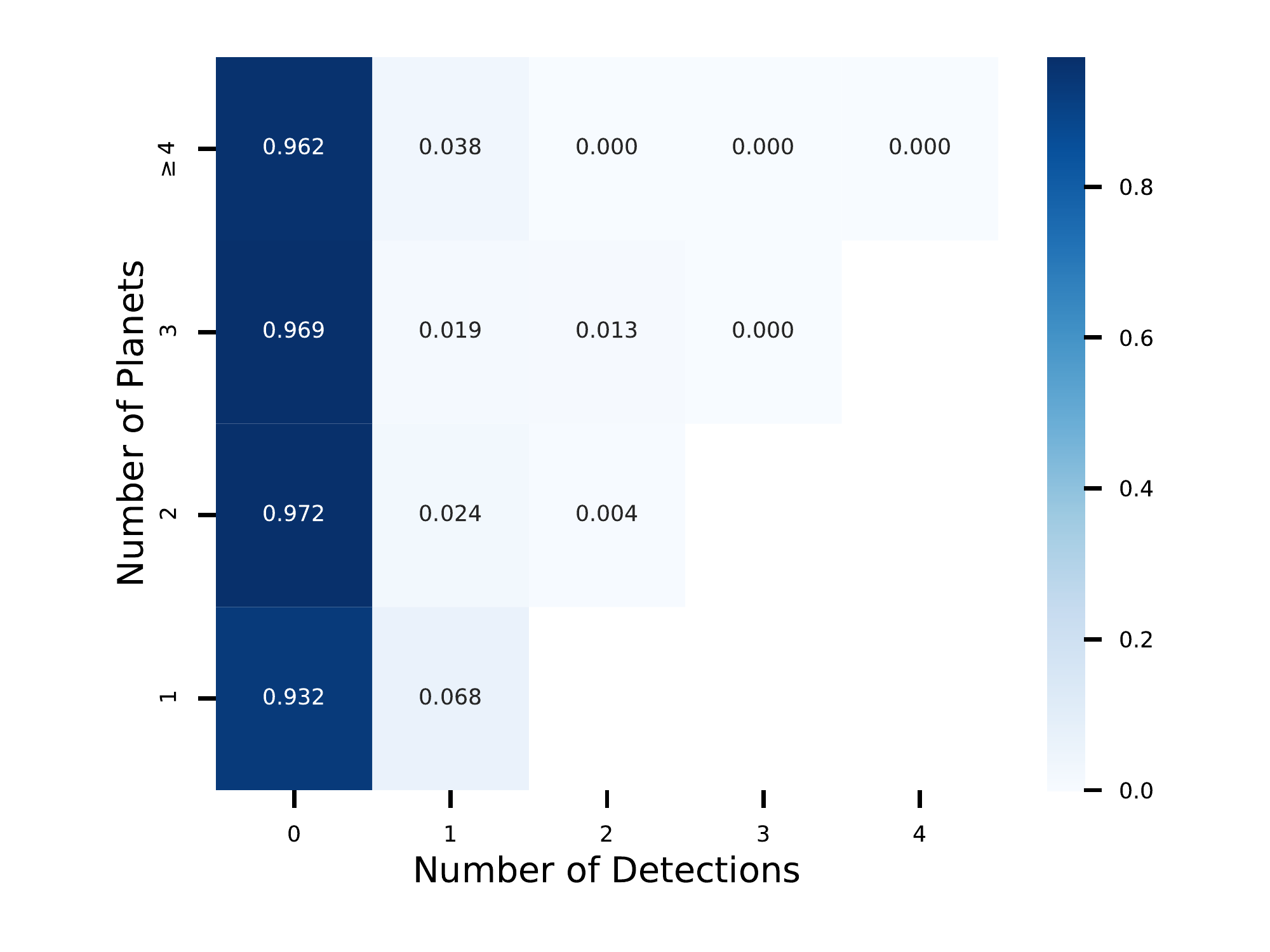}
  \end{minipage}
  \hfill
  \begin{minipage}[b]{0.48\textwidth}
    \includegraphics[width=\textwidth]{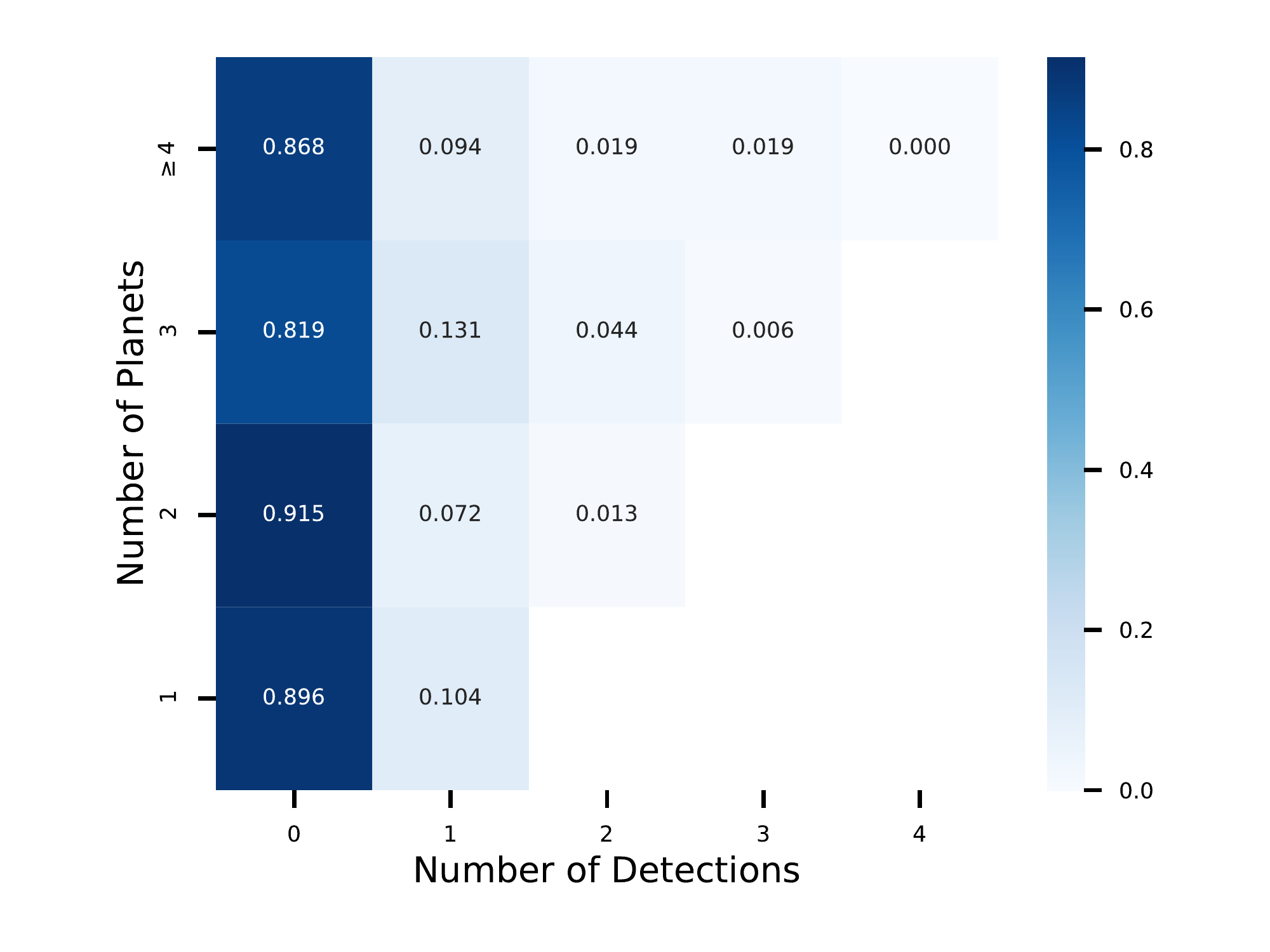}
  \end{minipage}
  \caption{(Left) The probabilities of detecting various scenarios with \tess\ in known \kep\ multi-planet systems for a detection threshold of $7.1\sigma$. For example, for a given two-planet system, the probability of detecting no planets in that system is 0.972, the probability of recovering one planet is 0.024, and the probability of recovering two planets is 0.004. (Right) The probabilities of detecting various scenarios in multi-planet systems for a detection threshold of $3\sigma$. Out of all the \kep\ multi-planet systems, \tess\ will have a very low probability of being able to recover more than one planet, regardless of what signal detection threshold is used.}
  \label{fig:multi}
\end{figure*}

Using the probabilities of detection for each planet, we next perform a simple analysis of how successful \tess\ will be at observing multi-planet systems.
We compute the probabilities for different scenarios (e.g. in a one planet system, the probability of detecting no planet or one planet) which is represented in Figure \ref{fig:multi}. 
Unfortunately, it is clear that for all multi-planet systems, \tess\ has a large probability of detecting none of the known transiting planets in a given system.
In only $\sim 5$ percent of known multiply-transiting systems do we predict more than one planet to be recovered, which agrees well with \cite{Sullivan15}'s finding that 5-10 percent of the KOIs will be recovered with \tess. 
The eventual launch of \plato, projected for the mid-2020s, will provide a better opportunity to investigate these systems.

\section{Improving Orbital Parameters of Dynamically Interacting Systems}
\subsection{KOI-142: A Test Case}
KOI-142 (Kepler-88) is a system with two known planets in near 2:1 resonance \citep{Nesvorny13}. 
KOI-142.01 is known for having one of the largest recorded TTV amplitudes of $\simeq$ 12 hours and is one of the only systems to show measurable transit duration variations (TDVs) \citep{Nesvorny13}. 
With this system we show an example of the expected improvement of planetary parameters by combining \tess\ and \kep\ data for well-characterizable systems.

KOI-142's parameters are already well-measured due to the uniqueness of fits from successful modeling. This success led \cite{Nesvorny13} to precisely determine the system's mass ratios. KOI-142's mass ratios were found to be: $M_b/ M_{\ast} < 5.2 \times 10^{-5}$ and $M_c/ M_{\ast} = (6.32)_{-0.13}^{+0.19} \times 10^{-4}$. 
With this information they determined that KOI-142b is a sub-Neptune class planet with a mass upper limit of 17.6 \mearth\ and that KOI-142c is a non-transiting planet with a mass of $215.9_{-7.5}^{+7.6}$ \mearth\ ($\simeq 0.7 M_{J}$).  
KOI-142b and KOI-142c orbit their central G-type star at periods of $\simeq$ 10.95 days and $\simeq$ 22.34 days, respectively.

Since we have precise values for many of this system's parameters, teams have been able to test theories for this system's formation and migration and have ruled out several possibilities due to the system's architecture \citep{Nesvorny13, Silburt15}. 

Nevertheless, there is still room for improvement in our understanding of this precisely characterized system and in systems that are not as well-measured.  We next describe our process for calculating the extent to which \tess\ will improve our measurements and understanding of KOI-142.

\subsection{MCMC Analysis}
\label{sec:mcmc}
Marginalized orbital parameters and uncertainties for KOI-142 from 14 quarters of \kep\ data were published by \citet{Nesvorny13}.
There is significant covariance between orbital parameters, as seen in those authors' Fig. 2 and Fig. 3. As those authors did not publish detailed posterior distributions of each parameter, the detailed orbits that are consistent with \kep\ data are not reproducible from that work. 
In addition, as three more quarters of \kep\ data were collected after the publication of that paper, we choose to re-fit the system entirely to make use of all available data. 
By re-fitting all transit times, as provided by \citet{Holczer16}, we can develop posterior distributions consistent with all public \kep\ data for KOI-142.

To find the posterior distribution and covariance between parameters, we perform a Markov Chain Monte Carlo (MCMC) analysis for this system.
We write a function that inputs potential sets of parameters into TTVFast \citep{Deck14} and outputs a series of transit times for all the planets in the system within a specified time interval.
We then compare all of the transits observed by \kep\ \citep{Holczer16} to the predicted times from TTVFast.
The \chisquare\ values from this comparison were outputted and used to create a likelihood function.
We combine our likelihood function with flat priors on all orbital parameters to perform an MCMC analysis for KOI-142 with thirteen degrees of freedom (the thirteen varying planetary parameters, listed in Table 3).
We use the affine invariant ensemble sampler \texttt{emcee} \citep{Goodman10, emcee}, using 150 walkers. We ran our MCMC 20,000 steps, discarding the first 10,000 steps as burn-in, and next use the Markov chains to compute the spread in transit times and uncertainties for future years when \tess\ will observe this system.    
We provide 30,000 samples from our posterior distributions as supplemental data associated with this manuscript.

\subsection{Spread in O-C Transits}
\label{sec:spreadTTV}
We use all of the Markov chains after burn-in and input each set of parameters into TTVFast, with the endpoint of integration set to the beginning of the year 2020.
We create an O-C (Observed minus Calculated) plot for the inner planet where Observed are the outputted transits from TTVFast and Calculated are the transit times based on a constant-period model \citep{Sterken05}.

\begin{figure*}[!tbh]
  \centering
  \begin{minipage}[b]{0.48\textwidth}
    \includegraphics[width=\textwidth]{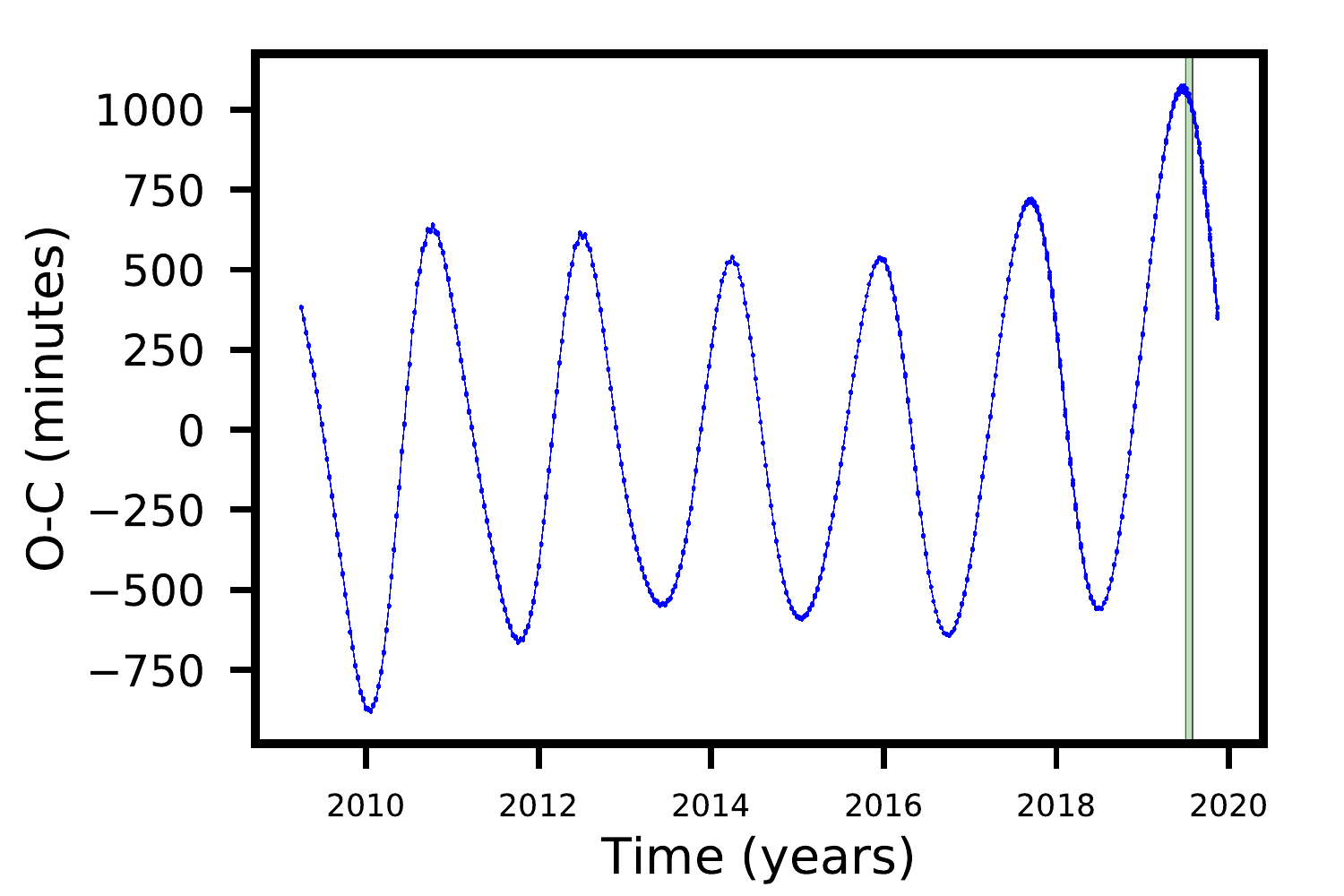}
  \end{minipage}
  \hfill
  \begin{minipage}[b]{0.48\textwidth}
    \includegraphics[width=\textwidth]{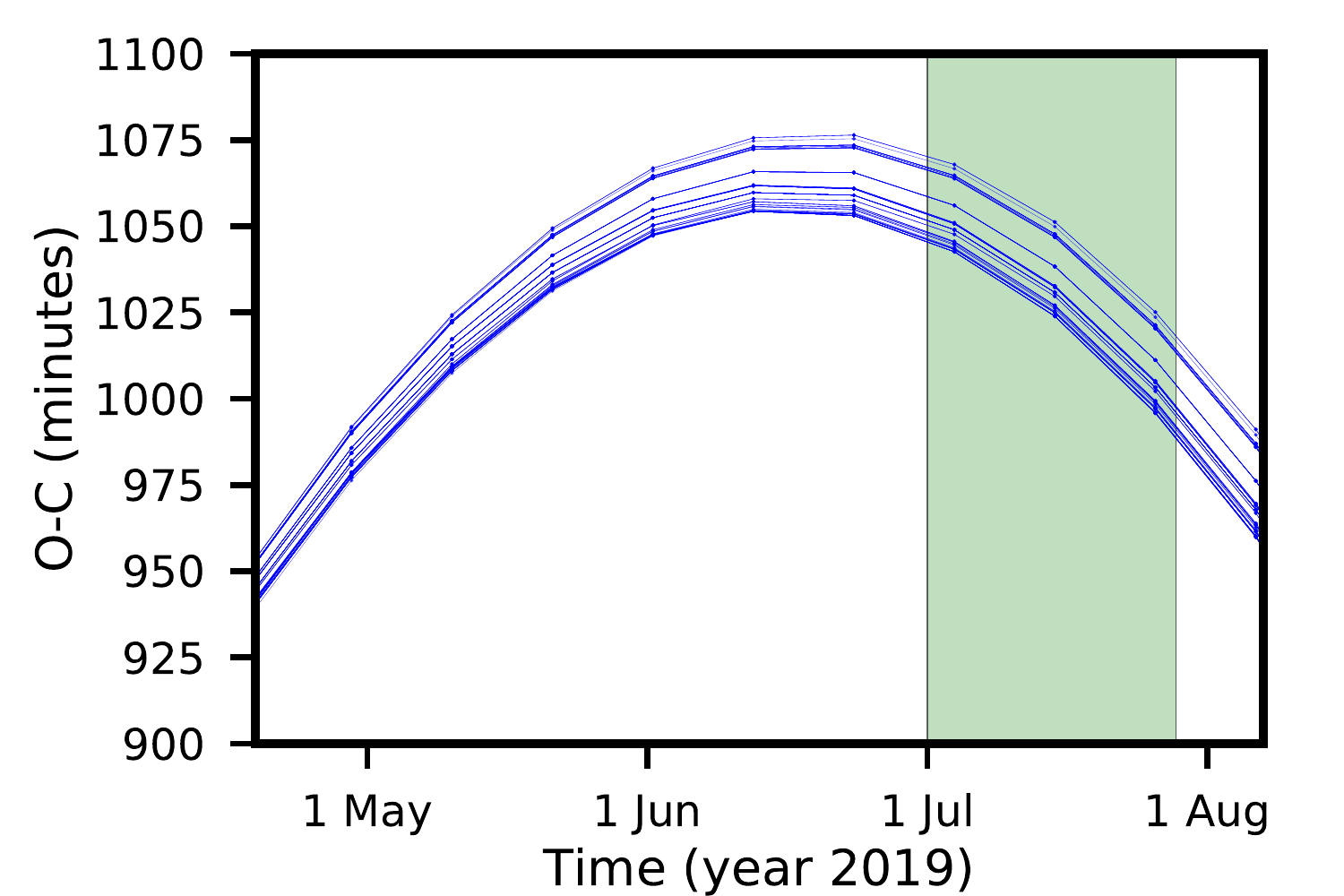}
  \end{minipage}
  \caption{(Left) O-C plot for KOI-142.01. The data begins in 2009 when \kep\ observes the system which continues through 2013. The data from 2013 onward was obtained by running all the possible sets of parameters from our Markov chains into TTVFast. (Right) A portion of the O-C plot in the year 2019, when \tess\ will observe KOI-142.  The obvious spread in transit times means that there is room for improvement in our measurements of KOI-142. The vertical green lines in both figures are representative of the time that \tess\ will observe KOI-142. Although the exact dates that \tess\ will observe this system are unknown, the width is representative of the actual expected duration of these observations.}
  \label{fig:OC}
\end{figure*}

The results from this section are in Figure \ref{fig:OC}. 
The leftmost figure shows KOI-142.01's TTVs in minutes from the year 2009 to 2020 using all of the Markov chains after burn-in.
The rightmost figure is a portion from this O-C plot in 2019, with the green band representing the time when \tess\ will observe KOI-142.01 (plotted using 100 randomly-selected sets of transits for clarity).
The large-amplitude resonant librations generate chaos in the system, as has been noted before for Kepler-36 \citep{Deck12}, so the future timing uncertainty grows non-linearly and faster than the expected linear rate given the uncertainty in the orbital period and number of missed transits (See also \citet{Mardling08}).

There is a very obvious spread in the transit times during the period that \tess\ will observe the system which means that there is certainly room for improvement in our knowledge of this system.
We then proceed to calculate the uncertainty in those transit times and determine by what factor we can reduce the transit uncertainty with \tess\ in July 2019. 

\subsection{Calculating Uncertainty in Transit Times}
\label{sec:uncertTTV}
\begin{figure*}[htbp!]
    \centerline{\includegraphics[width=0.8\textwidth]{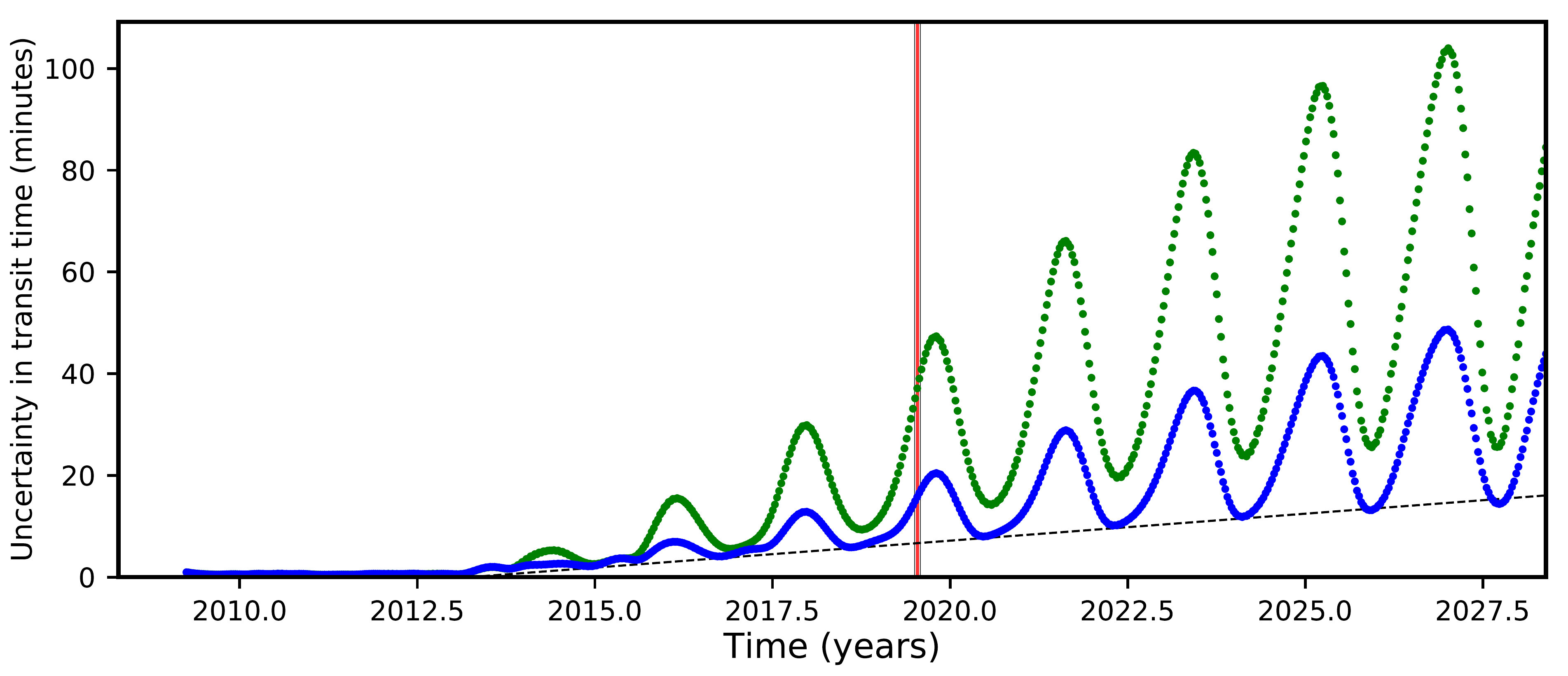}}
    \caption{(Top) The green points represent our uncertainty in transit times from the years 2009 to 2028. The blue points represent our new transit uncertainties when we include a theoretical \tess\ data point in July of 2019. The red vertical line depicts the month of July 2019 and the black dashed line illustrates the uncertainty in \kep\ data from propagating the uncertainty in period, assuming linear growth. From fitting three transits of simulated \tess\ data with the expected noise properties of that instrument, we expect the transit midpoint to be measured to a precision of 17 minutes, as described in Section 3.5. Models consistent with \kep\ data alone integrated to the \tess\ epoch have a spread in predicted transit time of 40 minutes. This means that KOI-142's transit times will be more precisely measured with \tess\ data in conjunction with \kep\ data rather than \kep\ data alone.}
    \label{fig:Uncert}
\end{figure*}

We use all of the Markov chains after burn-in and investigate the standard deviation of the expected time of each transit in the future, shown in green in 
Figure \ref{fig:Uncert}.

The uncertainty in the transits from 2009 until around 2013 are very small since \kep\ was observing KOI-142 during that time span. After 2013, there is an underlying, upward trend in transit uncertainties since we are no longer receiving data from the system. 

The behavior of the future TTVs is nonlinear and correlated with the TTVs shown in Figure \ref{fig:OC}. At some points in the TTV cycle, the spread in transit times is relatively large, and a precise observation of an observed transit at those times would be useful to provide better constraints on the orbital parameters of the system. 

In order to calculate \tess's transit uncertainty for KOI-142, we estimated the transit time uncertainty from TESS data, based on \tess's expected performance. 
For this, we develop simulated light curves using the \texttt{batman} package of \citep{Kreidberg15}.
We initially ran batman with our best fit parameters from Section \ref{sec:mcmc} at a time of inferior conjunction of $0.$, using a quadratic limb darkening model and limb darkening coefficients given by \cite{Claret17}'s models for a star with the physical parameters of KOI-142. 
We sample a flux value every 10 minutes (simulating three transits observed at 30 minute cadence) and add noise by sampling from a normal distribution with a standard deviation of 1500 ppm, \tess's expected sensitivity for KOI-142 \citep{Sullivan15}. 
We then treat those flux values with the added noise as ``observed'' data points.
We vary the transit time for our model 3000 times on a linear grid with a width of 0.2 days and calculate the \chisquare\ between each models' flux values and our noisy ``observed'' values.
We turn these \chisquare\ values into a posterior distribution on the time of transit.
The 16th and 84th percentile time values for our posterior were measured and we obtained a standard deviation on the expected time of transit of $0.0117$ days ($17$ minutes).

It is clear that with \tess's predicted transit uncertainty, we will be able to improve our measurements of this system.

\subsection{Improving our Measurements of KOI-142}
To quantify how much we can improve our measurements of the masses and eccentricities of these planets with \tess, we first went through a similar process as described in Section \ref{sec:mcmc}. 
The only difference is that we now include a new transit time observation in mid-July 2019 in our likelihood function. 
This new transit time observation is the mean of all of the transit times obtained from the Markov chains for mid-July 2019 ($BKJD = 3823.01$ days\footnote{BKJD = BJD - 2454833}) with an uncertainty given by the standard deviation calculated in Section \ref{sec:spreadTTV}. 
After running MCMC again, we analyze the best fit parameters and errors of the 16th, 50th, and 84th percentiles. 
Samples from this new posterior distribution are available on the journal website and in the source materials for this manuscript on the arXiv.

Table \ref{tab:params} lists each of the parameters values and associated errors for both the analysis without \tess\ data (Section \ref{sec:mcmc}) and with \tess\ data (this section).
The factor of improvement of our knowledge of this system with \tess\ data is in the last column. 

We expect the eccentricity measurements of both planets to be significantly improved, and the mass of the inner planet to be improved by more than 5 percent.
Even though the periods for both of these planets were well constrained with just \kep\ data, we expect to obtain even more precise period measurements with \tess, with the period uncertainties improving by 42 and 11\% for the inner and outer planet, respectively.
Due to the considerable improvement on most orbital parameters and planet-star mass ratios, the transit timing uncertainty is significantly reduced in the future: a single month of observations reduces the scatter in transit times expected in 2026 from 68 minutes to 33 minutes.

To illustrate this future improvement, we calculate the transit time uncertainty the same way as in Section \ref{sec:uncertTTV} but now also included our data associated with the new Markov chains.
In Figure \ref{fig:Uncert}, the green data points are transit uncertainties without any \tess\ data and the blue data points are transit uncertainties with our predicted \tess\ data in July 2019. 

There is clearly a large improvement that will be made when \tess\ observes this system; continued transit observations (such as through an extended \tess\ mission) will reduce the future uncertainties even further.
The best design for a \tess\ extended mission in regard to improving the transit uncertainties for \kep\ systems will take advantage of the oscillatory nature of Figure \ref{fig:Uncert}. 
As aforementioned, the behavior of the transit time uncertainty is due to the fact that at some points in the TTV cycle, the spread in the transit times is large.
If we are able to observe KOI-142 at the peaks in Figure \ref{fig:Uncert} where the uncertainty is largest, we will significantly improve our measurements of the system; conversely, observations where the uncertainty is small do not provide new information to distinguish possible physical models of the system.
For example, the theoretical \tess\ data point that we used in Section 3.5 to improve our measurements of KOI-142 was in July 2019, which is at a time very close to a large uncertainty peak in Figure \ref{fig:Uncert}. 
This concept can be applied to any \kep\ system in order to get the best results out of an extended mission.
The scatter in time generally follows the phase of the relative orientation of the two planets, which drives the slowly varying TTV signal.
We know that the period of a TTV signal obeys the super-period equation from \citep{Lithwick12}: 
\begin{equation}
    P^j = \frac{1}{|j/P' - (j-1) / P)|}
\end{equation}
where $P$ and $P'$ are the average orbital periods of
the inner and outer planet, and their resonance is $j:j-1$.
Since most observed TTV systems signals have a period ratio $P'/P$ within $1-3\%$ of commensurability, typical \kep\ systems observable with \tess, with periods of 10-20 days, have super-periods of 100-500 days.
Each super-period depends on the individual orbital parameters of that system. 
To measure any one system well, observations should be obtained at a particular phase with a cadence equal to the super-period. 
To maximize information about many systems, observations will need to be spaced to enable detection of sinusoidal signals at many different periods, so the ideal observations would not be clustered, but rather aperiodically scheduled to sample many different systems at their transit time uncertainty peaks.

In this work, we only consider the effects of TTVs, not transit duration variations (TDVs), which are observed for this system with a semiamplitude of $\sim 5$ minutes.
Our orbital parameters are similar to those of \citet{Nesvorny13} even without TDVs (but with three additional quarters of data which were not available at the time of that publication), suggesting the TDV information does not drive the fit. 
However, TDVs will be important for this system on \tess\ timescales. 
The authors of that paper note they cannot discriminate between two orbital solutions: one with $\Omega \approx 90^\circ$ and one with $\Omega \approx 270^\circ$. These two solutions imply a mutual inclination of $2.7^\circ$ or $4.5^\circ$, respectively, as the inner planet could have an inclination either just above or just below $90^\circ.$ However, these two models predict very different transit durations in 2019. In the case where $\Omega \approx 90^\circ$, the inner planet will have $b \approx 0$ when \tess\ observes the system, so the transit duration will be approximately 220 minutes. In the other case, $b \approx 0.8$ and the transit duration will be 140 minutes. These two will be easily separable with \tess.

For many dynamically interacting systems in which \tess\ will be able to detect
transits, the combination of \tess\ and \kep\ data will be useful to measure
system parameters better than \kep\ alone. A detailed analysis of many of these 
systems, originally characterized in \cite{Hadden17} is presented in \citet{Goldberg18}.

\begin{deluxetable*}{cccc}
\tablecaption{Mass and orbital parameter values found from our MCMC analysis for KOI-142. The middle columns are from the 50\% percentile of our posterior distribution, with 1-sigma errors. The right-most column is the percentage of improvement we have when using a theoretical \tess\ data point in 2019 in conjunction with \kep\ data. All parameters are defined at reference epoch BKJD (BJD-2454833) = 121.675215.}
\tablewidth{0pt}
\tablehead{
  \colhead{Parameter} &
  \colhead{Data without \tess} & 
  \colhead{Data with \tess} &
  \colhead{Improvement Factor (\%)} 
}
\startdata
$M_b$ [\msun] & $(2.106_{-0.668}^{+0.648}) \times 10^{-5}$ & $(2.145_{-0.637}^{+0.600}) \times 10^{-5}$ & 6.060 \\
$M_c$ [\msun] & $(6.252_{-0.056}^{+0.054}) \times 10^{-4}$ & $(6.246_{-0.052}^{+0.056}) \times 10^{-4}$ & 1.472 \\
\hline
$P_b$ [days] & $10.91655_{-0.00025}^{+0.00023}$ & $10.916569_{-0.00014}^{+0.00013}$ & 42.051 \\
$P_c$ [days] & $22.2650_{-0.0017}^{+0.0015}$ & $22.2649_{-0.0015}^{+0.0014}$ & 10.737 \\
\hline
$e_b$ & $0.05567_{-0.00021}^{+0.00021}$ & $0.05563_{-0.00018}^{+0.00019}$ & 9.819 \\
$e_c$ & $0.05767_{-0.00051}^{+0.00066}$ & $0.05762_{-0.00041}^{+0.00045}$ & 26.489 \\
\hline
$i_b\ [^o]$ & $88.417_{-0.195}^{+0.160}$ & $88.419_{-0.195}^{+0.157}$ & 0.714 \\
$i_c\ [^o]$ & $86.361_{-0.707}^{+0.624}$ & $86.472_{-0.830}^{+0.559}$ & -4.396 \\
\hline
$\omega_b\ [^o]$ & $179.178_{-1.355}^{+1.337}$ & $179.002_{-0.919}^{+0.891}$ & 32.762 \\
$\omega_c\ [^o]$ & $1.714_{-0.818}^{+1.044}$ & $1.606_{-0.601}^{+0.809}$ & 24.314 \\
\hline
$\Omega_b\ [^o]$ & --- & --- & --- \\
$\Omega_c\ [^o]$ & $359.61_{-1.75}^{+1.79}$ & $359.36_{-1.31}^{+1.34}$ & 25.078 \\
\hline
$\lambda_b\ [^o]$ & $263.85_{-1.30}^{+1.32}$ & $264.02_{-0.85}^{+0.88}$ & 33.873 \\
$\lambda_c\ [^o]$ & $335.80_{-1.25}^{+1.20}$ & $335.96_{-1.01}^{+0.89}$ & 22.186 \\
\hline
$\sigma_{2019.5}$ [min.] & 34.61 & 14.73 & 57.440 \\
\hline
$\sigma_{2026.5}$ [min.] & 67.34 & 32.92 & 51.114 \\
\hline
\enddata
\tablenotetext{}{}
\label{tab:params}
\end{deluxetable*}

\section{Tidal Decay of Hot Jupiters}
Hot Jupiters have been detected by numerous ground and space-based observations due to their large sizes and short periods, making them easy to find in both RV and transit surveys \citep{Mayor95, Bakos04, Mccullough2005, Mccullough2006, Pollacco06, Brahm16}.
Hot Jupiters are still interesting targets, with many questions about the formation and interior structure of these planets still outstanding \citep[e.g.][]{Guillot05, Guillot06, Nelson17, Dawson18} 

\citet{Birkby14} searched for evidence of tidal decay in the population of then-known hot Jupiters, finding inconclusive results from the available data.
\cite{Patra17} investigated the transit timing anomaly of WASP-12b, a hot Jupiter with an orbital period of 1.09 days, and modeled this planet's orbital period, arguing that WASP-12b is more likely in orbital decay than in a precession cycle. 
\cite{Patra17} acknowledge, however, that more observations are necessary to completely rule out the precession model.
Further investigations of WASP-12b's transit timing anomaly agree that WASP-12b is likely in orbital decay and that classifying WASP-12 as a subgiant accurately explains the observed change in period as well as its decay timescale of 3 Myr \citep{Weinberg17, Bailey18}.

\cite{Ragozzine09} demonstrated how measuring apsidal precession enables one to infer properties of the interior structure of the planet and star, such as \q. 
\q\ is the tidal quality parameter of a star which is the measure of the star's response to tidal distortion due to a perturbing body. 
\cite{Ragozzine09} similarly mention that a longer baseline will be necessary in order to measure apsidal precession in many very hot Jupiter systems. 

In Section \ref{sec:sensitivity}, we showed that \tess\ will be able to observe many of the hot Jupiters detected by \kep.
We now investigate whether \tess\ will be powerful enough and will considerably extend the baseline of observations such that we can measure orbital decay or precession and learn more about the interior structure and formation of hot Jupiter planets and their host stars.

\subsection{Hot Jupiter Transit Uncertainties in July 2019 with \kep\ and \tess}
\label{sec:kep_tess_uncert}

In the following analysis we consider the 366 planets from Section \ref{sec:sensitivity} with radii smaller than 30 \rearth\ that were detectable in two sectors at a $3\sigma$ level.

We propagate the errors from the period and transit epoch (obtained from the NASA Exoplanet Archive cumulative KOI table) into July 2019 to compute the uncertainty in transit times given \kep\ data alone. 

To compute the \tess\ uncertainty we simulate light curves using the \texttt{batman} package of \citet{Kreidberg15} as in Section \ref{sec:spreadTTV}.
We create a model for each planet and add noise by sampling from a normal distribution with a standard deviation give by the expected noise from \tess\ calculated in Section \ref{sec:methods} and listed in Table \ref{tab:tess_sensitivity}.
We contaminate the flux values of the model depending on the \tess\ contamination ratio for that system and use these new flux values as ``observed'' data. 
We vary the time of inferior conjunction many times and use the same limb darkening coefficients as in Section \ref{sec:spreadTTV}. 
This time, we sample a flux value every 2 minutes within a 30 minute interval, assuming most detectable hot Jupiters will be observed at short cadence, but we stack transits together depending on the planet's period so the time of sampling varies for each system.
We find the posterior distribution of transit times, measure the $1\sigma$ width of this distribution, and take that as the expected transit timing uncertainty for each system.

To find the total uncertainty in July 2019 we combine the uncertainty in the transit times inferred from \tess\ data with the uncertainty from \kep\ data.

\subsection{Hot Jupiter Transit Times in July 2019}
\label{sec:transit_times}

We first calculate the expected transit time in July 2019 for these systems based on a constant-period model. 

We next calculate the transit time in July 2019 for these systems using an orbital-decay model.
To do this, we first find the change in period due to orbital decay using \cite{Patra17}'s Eqn. (14): 
\begin{equation}
\frac{dP}{dt} = -\frac{27\pi}{2Q_{\ast}}\left(\frac{M_{p}}{M_{\star}}\right)\left(\frac{R_{\star}}{a}\right)^5,
\end{equation}
where \q\ is the tidal quality parameter of the star. 
Since we do not have measurements for $M_p$ for many systems, we first assume all planets with radii larger than 8 \rearth\ to be a Jupiter mass planet.
We calculate $M_p$ for planets with radii less than 8 \rearth\ by using the planet mass-radius relationship of \cite{Lissauer11}. 
Given the predicted period for a linear ephemeris and our given uncertainties in the transit time, we determine a threshold value for \q\ that will make each system detectable at $3\sigma$.

\subsection{Candidate Decaying Systems}
We compute the ratio of the transit difference between the models and the total uncertainty in Sections \ref{sec:kep_tess_uncert} and \ref{sec:transit_times}. 
We calculate the threshold \q\ values that would make orbital decay detectable at $3\sigma$, which we call \qcrit. 
For every system, \qcrit\ is the value at which we expect to detect tidal decay if the true \q\ for that star is less than \qcrit. 
Therefore, larger values of \qcrit imply a higher likelihood of detection of tidal decay for given stellar parameters.

\begin{figure}[htbp!]
\centerline{\includegraphics[width=0.5\textwidth]{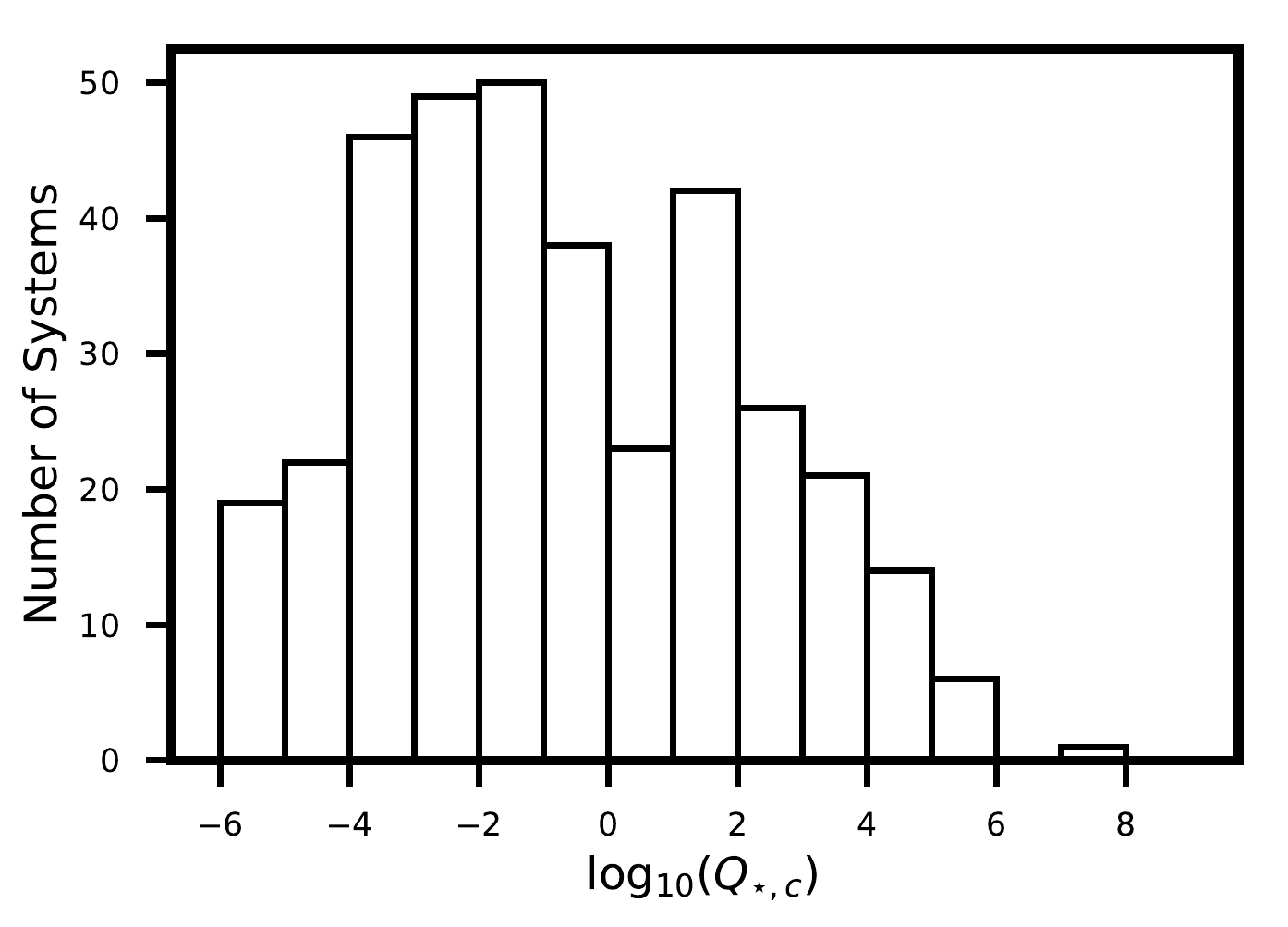}}
\caption{Histogram showing the spread in \qcrit\ values (calculated using planet mass-radius relation) for systems that have a greater than 50\% chance of detection with \tess\ in two sectors at the $3\sigma$ level. \ The seven systems with \qcrit\ greater than $10^5$ will be good candidates for potentially detecting orbital decay, or placing meaningful limits on the values of \q.
}
\label{fig:HJ_hist}
\end{figure}

Figure \ref{fig:HJ_hist} shows the spread in \qcrit\ values that would make orbital decay detectable at $3\sigma$. 
A typical value of \q\ is around $1\times10^5$ \citep{Dobbs-Dixon04} and so any planets that have \qcrit\ $\geq 1\times10^5$ will be good candidates to potentially measure orbital decay.

We find two confirmed systems and six candidates that have \qcrit\ $\geq 1\times10^5$ and are thus good candidates for orbital decay detection. 
We inspect the light curves of each candidate system by eye. We conclude that KOI-7430 is a likely false positive and exclude it from our analysis, leaving seven systems in total.
Two of these are confirmed planets: Kepler-2\,b and Kepler-13\,b (KOI-2 and KOI-13).
The other five are planet candidates; these are KOI-1075, 3156, 5220, 5353, and 7259.

In Table \ref{tab:HJ} we list the \qcrit\ values for ten detectable planets at the $3\sigma$ level in two sectors. 
We calculated \qcrit\ as discussed in the previous sections (using the planet mass-radius relationship to compute masses for planets smaller than 8 \rearth).
As the planet masses are typically unknown, we also provide values for $Q/M_p$, in units of Jovian masses. As masses of these planets are measured, these updated masses can be used directly to update the expected \qcrit\ values.

\tess\ is essential for this task. 
For each of the seven candidate or confirmed systems with $\log$(\qcrit) $> 5.0$, we repeat this exercise considering \kep\ data alone, finding that in all cases \tess\ transits provide a significant extra constraint. 
For the median system, we find $\log$(\qcrit) increases by 0.52 when \tess\ observations are added to the existing \kep\ data, increasing our sensitivity to \q\ by a factor of 3.1.

\begin{deluxetable*}{lcccccccc}
\tablecaption{List of parameter and critical \q\ values for all detectable systems at the $3\sigma$ level in two sectors. These systems are ordered by KOI name.}
\tablewidth{0pt}
\tablehead{
  \colhead{KOI} &
  \colhead{\kep\ Name} & 
  \colhead{Period} &
  \colhead{Radius} &
  \colhead{\mstar} &
  \colhead{$\sigma$ \kep} &
  \colhead{$\sigma$ \tess} &
  \colhead{$\log_{10}$(\q$_{,c}$)\tablenotemark{1}} &
  \colhead{$\log_{10}$(\q$_{,c} / M_J$)\tablenotemark{2}} \\
  \colhead{} &
  \colhead{} & 
  \colhead{(days)} & 
  \colhead{(\rearth)} & 
  \colhead{(\msun)} &
  \colhead{(min.)} &
  \colhead{(min.)} &
  \colhead{} & 
  \colhead{}
}
\startdata
K00001.01 & Kepler-1\,b & 2.470613 & 13.04 & 0.97 & 0.06 & 0.16 & 4.03 & 4.03 \\
K00002.01 & Kepler-2\,b & 2.204735 & 16.10 & 1.45 & 0.11 & 0.28 & 5.06 & 5.06 \\
K00003.01 & Kepler-3\,b & 4.887803 & 4.82 & 0.83 & 0.52 & 0.26 & 0.79 & 1.88 \\
K00004.01 & & 3.849372 & 12.94 & 1.48 & 3.31 & 2.68 & 3.76 & 3.76 \\
K00007.01 & Kepler-4\,b & 3.213669 & 4.13 & 1.10 & 1.86 & 5.75 & 1.63 & 2.86 \\
K00010.01 & Kepler-8\,b & 3.522498 & 14.59 & 1.13 & 0.31 & 1.37 & 3.15 & 3.15 \\
K00012.01 & Kepler-448\,b & 17.855222 & 13.16 & 1.39 & 1.12 & 1.86 & -0.46 & -0.46 \\
K00013.01 & Kepler-13\,b & 1.763588 & 21.42 & 2.47 & 0.13 & 0.30 & 5.79 & 5.79 \\
K00017.01 & Kepler-6\,b & 3.234699 & 13.06 & 1.05 & 0.20 & 1.06 & 3.27 & 3.27 \\
K00018.01 & Kepler-5\,b & 3.548465 & 14.92 & 1.32 & 0.28 & 1.51 & 3.32 & 3.32 \\
\enddata
\tablenotetext{1}{Calculated using planet masses obtained from the mass-radius relation for planets with radii $<$ 8 \rearth. Otherwise, assumed planets with radii $\geq$ 8 \rearth\ to be Jupiter mass planets.}
\tablenotetext{2}{Leaving planet masses as an input parameter}
\tablecomments{This table will be published in its entirety in machine-readable format on the journal website. 
A portion is reproduced here as a guide for formatting. A version is also available in the source materials
for this manuscript on the arXiv.}
\label{tab:HJ}
\end{deluxetable*}

\subsection{Fisher matrix analysis} 
\label{sec:fisher}
One may ask how new data from \tess\ can compete with the highly-precise four-year \kep\ dataset. Or, if \kep\ was not sensitive to tidal decay for the set of planets it observed, how will \tess\ be sensitive to it. The answer is that the phase change of a decaying planet is quadratic, so the longer one waits, the more powerful the lever arm gets for constraining this curvature. We perform a Fischer-matrix analysis to determine the minimum-variance bound for each of three parameters, $\mathbf{a} = (T_0, P,\dot{P})$, in which the model time is: \begin{equation}
t_m(i) = T_0 + i P + \onehalf i^2 \dot{P},
\end{equation}
where $i$ is the transit number with $0$ at the center of the \kep\ dataset. 
The figure-of-merit is 
\begin{equation}
\chi^2 =  \sum \frac{(t(i)-t_m(i))^2}{\sigma_i^2}
\end{equation}
where the measured mid-time and uncertainty of a measured transit $i$ are $t(i)$ and $\sigma_i$, respectively. The curvature of the $\chi^2$ surface informs the minimum size of uncertainties in the parameters, and this calculation is shown in the Appendix. 
The uncertainties on the three parameters from \kep\ alone are:
\begin{eqnarray}
\sigma_{T0} &=& 3/2 (\sigma/\sqrt{N}) \\
\sigma_P &=& 2 \sqrt{3} (P/T)(\sigma/\sqrt{N}) \\
\sigma_{\dot P} &=& 12 \sqrt{5} (P/T)^2 (\sigma/\sqrt{N}),
\end{eqnarray}

where $T_0$ is the measured time of transit, $P$ the orbital period, $T$ the length of the survey, and $N$ the number of observed transits.
Note that if the survey is of high duty cycle, which is true of the \kep~survey, $N \simeq T/P$. Therefore the precision on period determination improves as the $3/2$ power of survey length  $T$, and the precision on period change rate improves as the $5/2$ power of survey length.

We then consider a second dataset augmenting the first dataset, which has the centers of the two datasets offset by $D$ in time, with a certain value of $\sigma_2/\sqrt{N_2}$, and a time baseline $T_2$ (within the second dataset, over which transits are uniformly spread).  If the values of these survey parameters are as given in Model 2 of Table~5, for instance, with addition of new data the uncertainty improves by 22\% for $T_0$, worsens by 1\% on $P$, and improves by 48\% for $\dot P$. We suppose the formal uncertainty of $P$ can worsen because of correlations between $P$ and the other parameters when considering a heterogeneous dataset. The period derivative itself, the quantity of interest for tidal decay, always improves with more data, however. 

The uncertainty in the period derivative, the quantity of interest for measuring orbital tidal decay, decreases as the square of the observed time baseline, and the square root of the number of observed transits as shown in Equation 6. Therefore, for the purposes of maximizing the number of detections of tidal dissipation in an extended mission, additional campaigns focused on the \kep\ field should be scheduled at the end of the extended mission. This scheduling would allow for the largest possible change in period between the start of the \kep\ mission and the final \tess\ observations.

Two real examples, to which we apply this formalism, are KOI-13 and KOI-18. The former is systematics-dominated in \kep, and transit times uncertainties determined by \tess\ may be only $\sim 2.3$ times larger than \kep. The uncertainty on period change will decrease by a factor of about $3$, with 1 sector of \tess\ observations.  In the case of KOI-18, both \kep\ and \tess\ are photon-limited, and hence a more modest improvement is expected (Table~5).

\begin{deluxetable*}{l|ccc|ccc}
\tablecaption{Factor by which the uncertainty in a quadratic timing model changes, when an additional dataset is added to the \kep\ data.  Parameters of the survey are explained in the text of Section~\ref{sec:fisher}.}.
\tablewidth{0pt}
\tablehead{
  \colhead{ } &
  \colhead{ new } & 
  \colhead{ survey } &
  \colhead{ parameters } &
  \colhead{ }&
  \colhead{ result} &
  \colhead{ }\\
  \colhead{ } &
  \colhead{ $\sigma_2 / \sqrt{N_2}$} & 
  \colhead{ $T_2$} &
  \colhead{ $D$} &
  \colhead{ $\sigma_{T0}$}&
  \colhead{ $\sigma_{P}$}&
  \colhead{ $\sigma_{\dot P}$}\\
  \colhead{ Model } &
  \colhead{ $(\sigma / \sqrt{N})$} & 
  \colhead{ (yr)} &
  \colhead{ (yr)} &
  \colhead{ (old value)}&
  \colhead{ (old value)}&
  \colhead{ (old value)} 
}
\startdata
1 & $100$ &  0.08 &  $10.05$ & $0.887$  & $1.013$ &  $0.770$ \\
2 & $50$ &  0.16 &  $10.05$ & $0.780$  & $1.013$ &  $0.520$  \\
KOI-13 & $17.0$ & $0.074$ & $8.25$ &  $0.714$ & $0.979$&     $0.315$ \\
KOI-18 & $41.2$ & $0.074$& $8.25$ & $0.798$  & $1.009$ &   $0.599$
\enddata
\label{tab:fisher}
\end{deluxetable*}

\section{Discussion and Future Prospects}

\subsection{The Noise Properties of \tess\ Data}

We analyzed \tess's capabilities in detecting and improving our measurements of planetary systems in the \kep\ field. 
By converting data from the NASA Exoplanet Archive into probabilities of detection for each planet, we found 260 (161) planets in one sector of observations and an additional 120 (18) planets in two sectors that have a strong chance of being recovered at the $3\sigma$ ($7.1\sigma$) level. 
The majority of these recovered signals are hot Jupiter planets; nevertheless, there are still some smaller, rocky planets that can be recovered, such as KOI-6635.01, which orbits a small, bright star.

Although the \kep\ signal detection threshold was $7.1\sigma$, we believe that a $3\sigma$ level will be largely sufficient for \tess\ in the \kep\ field since the goal is to characterize planets that are already known to exist, rather than detecting new planets. 
Additionally, it is important to note that throughout this analysis we made the following assumptions: 1) we assumed that \cite{Sullivan15}'s model is correct in predicting the total noise \tess\ will experience while observing each system, 2) that the total noise was completely white, and 3) that the contamination ratios from \cite{Brown11} and \cite{Stassun17} are correct. 
If any of these assumptions are incorrect or too restrictive, our results will differ from what we originally predicted. 
With the public availability of \tess\ data and pipelines to produce light curves \citep[e.g.][]{Feinstein19}, the noise properties of stars hosting known planets can be understood (T. Daylan et al. in prep).

The expected yield of planets in the \kep\ field will provide a direct opportunity to characterize the noise properties and general performance of the \tess\ detector.
For every confirmed planet or planet candidate in the \kep\ field, we provide a probabilistic forecast of detection in \tess\ data. 
By comparing these results to the actually-detected planets in late 2019, we will be able to understand specific weaknesses in the assumptions used in the development of the \tess\ Input Catalog \citep[TIC,][]{Stassun17}, which is used for target selection in the primary mission and will likely be used in the same way in an extended mission.
For example, if planet yield in regions of high stellar density is higher than expected, it might suggest that the assumptions in the TIC about stellar contamination are too conservative. 
The precision of \tess\ is not higher than \kep. In \kep\ data, the
measurement of $\eta_\oplus$ was challenging because stellar variability was higher than expected. However, given the existence of \kep\ data, we already know the intrinsic variability of stars at the level of \tess\ precision, so any discrepancies in planet yield are, for \tess, more likely due to limitations of the instrument or the input catalog, rather than in inherent variability of stars which is now better understood.
As the \kep\ field provides the largest sample of known transiting planets available, these predictions provide the best available opportunity to understand the performance of \tess\ on known, characterized planetary signals.

\subsection{The Future of Multi-Transiting Systems}

Using our probabilities of detection for each planet, we organized our data such that we could determine how capable \tess\ will be in detecting multi-planet systems.
We found that \tess\ will be expected to recover more than one planet in only $\sim 5$ percent of known multi-planet systems.
Future studies dealing with multi-planet systems will likely be more successful with \plato\ observations of the \kep\ field, perhaps in the mid-2020s.

For planets that \tess\ will be able to detect, we expect that \tess\ will be a very useful tool in improving our measurements of these systems.
For KOI-142, we predict that \tess\ will improve a majority of the planetary parameters as well as star-planet mass ratios, which yields an improvement of over 50\% in the transit timing uncertainties in the future.

The combination of \kep\ and \tess\ will also provide a useful test case for a possible extended \tess\ mission. The primary \tess\ mission will cover $\approx 80\%$ of the sky. An extended mission proposal is now submitted; additional extensions are still subject to change. 
With more data, smaller planets, which are more often found in multiple-planet systems, can be recovered. Some of these multiple-planet systems will exhibit TTVs and TDVs; by combining these observations together across multiple sectors with large data gaps, masses and orbital parameters will be measurable for systems without the need for any additional follow-up resources. Observations of the \kep\ and \tess\ fields taken together will provide insights about best practices to combine these data sets to look for dynamical effects.

\subsection{Characterizing Tidal Dissipation}

Since we predict \tess\ will be most sensitive to hot Jupiters, we analyzed whether \tess\ will be able to detect tidal orbital decay in hot-Jupiter systems. 
We found two confirmed and five candidate planets that will be good candidates for detecting orbital decay. 
If we are able to detect orbital decay in any of the systems in the \kep\ field we will be able to better understand their interiors and perhaps more accurately test theories of planetary formation and migration.

One may ask why \tess\ is strictly necessary for this task, as ground-based
observations should be achievable for these giant planets. 
In this work, to measure a discrepancy in transit times in 2019 from the previous linear ephemeris, we assumed that all transits over 27 days could be observed. For the cases of Kepler-2 and Kepler-13, this is 12 and 15 transits, respectively.
While 15 transits could be observed from the ground, that would require a significant investment of observing resources spread over months to collect the same amount of data for a single target. Additionally, these data may lead to a lower precision than what can be acquired from \tess\ as ground-based data contains time-correlated noise due to atmospheric variability on few-minute timescales, which can significantly inhibit precise transit time measurements \citep[e.g.][]{Pont06}.
Therefore, \tess\ data provide the best opportunity to combine new observations with \kep\ data to measure tidal orbital decay of hot Jupiters, although these data could certainly be combined with additional ground- or space-based photometry once candidates are identified.

\subsection{Planetary Evolution with \tess}

We find in general that long time baseline observations of planetary systems with space-based observatories can be useful for
understanding the physical parameters and long-term evolution of planetary systems. 
This is applicable to combinations of \kep\ and \tess\ data, but also to missions like \textit{K2}, where data spanning multiple years when multiple campaigns overlap can be used to confirm and measure masses of dynamically interacting planetary systems (A. Hamann et al. in prep).

In time, as \tess\ continues to re-observe the \kep\ field through an extended mission, each additional campaign will yield approximately 50 more planets smaller than the size of Neptune that have a good chance of being detected (Table 1). 
Since these smaller planets are typically found in multi-planet systems, we will be able to better characterize the systems through TTVs and TDVs. These observations will enable us to improve our understanding of transiting and non-transiting planets in these systems. In some cases, we might observe planet precession through duration variations due to a non-transiting perturber, identifying new planets in these systems that were previously missed \citep{Ribas08, Mills17}. Moreover, most sub-Neptunes in multiple-planet systems are confirmed planets \citep{Morton16}, so these observations will largely probe the dynamical evolution of bona fide planets.

Although \tess\ may not be as sensitive to as many planets as \kep, we show that \tess\ will be extremely effective in improving our measurements and understanding of certain systems.  This improvement will only be enhanced by an extended mission that continues to include observations of the \kep\ field, enabling the detection of smaller confirmed planets.

\subsection{Possible \tess\ Extended Mission Strategies}

A \tess\ extended mission will have many disparate goals. Here, we outline possible strategies to maximize the scientific yield of already-known planets in the \kep\ field.
To re-detect as many planets as possible, especially the small planets that are commonly found in multiple-planet systems \citep[e.g][]{Fabrycky14}, maximizing the number of sectors in which the \kep\ field is re-observed is the primary requirement: observations in 6 or more sectors is required to detect even 10\% of these systems.

Many of these systems that would be detected through additional sectors of observations have transit timing variations. The ideal strategy for maximizing our ability to characterize these systems would be aperiodic observations of the \kep\ field. Precise transit times are needed to constrain orbital parameters like massses and eccentricities, but as shown in Section 3, the uncertainty in future transit times is a strong function of relative orbital phase for these systems. Therefore, observations at particular phases in the ``super-period'' are required. As every system has a different super period, typically over the range 100-500 days, to maximize the power of \tess\ to characterize these systems we require sectors spaced in time to provide power over as much of this range as possible.

Finally, to measure tidal dissipation, as shown in Section 4, observations should be scheduled as late in the mission as possible. This is intuitive: as in this case the planet's orbital period is monotonically decreasing, at the end of the mission the period has the largest change in period from the start of the mission. This case is somewhat at odds with the previous case. However, as this one is likely to provide the largest scientific yield from the mission (at present, there is only one system with tentative tidal dissipation \citep{Patra17}), we encourage the \tess\ team to consider the viability of this strategy in any and all extended mission plans to guarantee the community can get the most out of an extended mission covering the \kep\ field.
We are looking forward to receiving data from \tess\ in the \kep\ field and gaining a more comprehensive understanding of planets beyond our solar system.

\acknowledgments

We thank Matthew Payne (CfA | Harvard and Smithsonian) for helpful comments which improved the quality of this manuscript and Adina Feinstein (Chicago) for her assistance with the FITS standard. We thank the anonymous referee for their rapid response and careful consideration of this manuscript.

C.N.C. thanks the University of Chicago College Research Fellows Fund for their generous support.
Work by B.T.M. was performed under contract with the Jet Propulsion
Laboratory (JPL) funded by NASA through the Sagan Fellowship Program executed
by the NASA Exoplanet Science Institute.

This paper includes data collected by the \kep\ mission. Funding for the
\kep\ mission is provided by the NASA Science Mission directorate.
We are grateful to the entire \kep\ team, past and present.
These data were obtained from the Mikulski Archive for Space Telescopes
(MAST).
STScI is operated by the Association of Universities for Research in
Astronomy, Inc., under NASA contract NAS5-26555.
Support for MAST is provided by the NASA Office of Space Science via grant
NNX13AC07G and by other grants and contracts.

This research has made use of the NASA Exoplanet Archive, which is operated by the California Institute of Technology, under contract with the National Aeronautics and Space Administration under the Exoplanet Exploration Program

\software{%
  numpy \citep{numpy}, matplotlib \citep{Hunter07}, TTVFast \citep{Deck14}, emcee \citep{emcee}, batman \citep{Kreidberg15}, pandas \citep{McKinney10}, seaborn \citep{seaborn}
  }

\facilities{\kep, Exoplanet Archive}
 
\appendix
\section{Derivation of Fisher Matrix}
To perform a Fisher matrix analysis, we form the array: 
\begin{equation}
\frac{\partial^2 \chi^2}{\partial a_{j} \partial a_{k}},
\end{equation}
from which the correlation matrix can be derived as 
\begin{equation}
C_{j,k} = \bigg( \frac{1}{2} \frac{\partial^2 \chi^2}{\partial a_{j} \partial_{k}} \bigg)^{-1}.
\end{equation}

The analytic derivatives of the $\chi^2$ function are: 
\begin{equation}
\frac{\partial^2 \chi^2}{\partial a_{j} \partial a_{k}} = \begin{bmatrix}
   2 & \sum i & \sum i^2  \\
   \sum i & 2\sum i^2 & \sum i^3 \\
   \sum i^2 &  \sum i^3 & \sum i^4/2
   \end{bmatrix}, \label{eqn:Lcurvature}
\end{equation}
where the sums are over the $N$ data points of $i$.
 Evaluating these, since the origin of the $i$ array is at the center of the data, odd functions of $i$ cancel to zero, whereas even functions can be integrated. The result is 
 \begin{equation}
\frac{\partial^2 \chi^2}{\partial a_{j} \partial a_{k}} = N/\sigma^2 \begin{bmatrix}
   2 & 0 & (T/P)^2/12  \\
   0 & (T/P)^2/6 & 0 \\
   (T/P)^2/12 &  0 & (T/P)^4/160
   \end{bmatrix}.
\end{equation}

We evaluate this array first for the \kep\ data alone, assuming that $N$ transit timings are taken uniformly spread through the timespan of $T=4.02$~yr, and that each transit time has an equal uncertainty of $\sigma$.  We find: 
\begin{equation}
C_{j,k} = \sigma^2/N \begin{bmatrix}
   9/4 & 0 & -30 (P/T)^2 \\
   0 & 12 (P/T)^2 & 0 \\
   -30 (P/T)^2 &  0 & 720 (P/T)^4 
   \end{bmatrix}. \label{eqn:cjk}
\end{equation}
For combining \tess\ data with \kep\ data, we evaluate equation~\ref{eqn:Lcurvature} numerically, then invert, giving the values in Table 5.

\end{document}